\documentclass[12pt,draftcls,onecolumn]{IEEEtran}
\usepackage{graphicx}
\usepackage{psfrag}
\usepackage{amssymb}
\def\be{\begin{equation}}
\def\ee{\end{equation}}
\def\bea{\begin{eqnarray}}
\def\eea{\end{eqnarray}}
\def\beann{\begin{eqnarray*}}
\def\eeann{\end{eqnarray*}}

\def\ns{\hspace{-1mm}}

\def\gA{{\cal A}}

\def\gR{{\cal R}}

\def\gV{{\cal V}}

\begin{document}
%
%\title{
%A Direct Proof of Theorem 1 in ``Structural Invariant Subspaces of Singular Hamiltonian Systems
%and Nonrecursive Solutions of Finite-Horizon Optimal Control Problems''}
\title{\Large
A Note on the Dimensions of the Structural Invariant Subspaces of the Discrete-Time Singular Hamiltonian Systems}
\author{G.~Marro\\[2 mm]
{\small  Dipartimento di Elettronica, Informatica e Sistemistica,
Universit\`a di Bologna}\\
{\small Viale Risorgimento 2, 40136 Bologna - Italy}\\
{\small E-mail: \tt giovanni.marro@unibo.it}\\[2mm]
}
\date{}
\maketitle
\thispagestyle{empty}
%%%%%%%%%%%%%%%%%%%%%%%%%%%%%%%%%%%%%%%%%%%%%%%%%%%%%%%%%%%%%%%%%%%%%%%%%%%%%%%%
%
\begin{abstract}
\noindent
The structural invariant subspaces of the discrete-time singular Hamiltonian system are used in  \cite{TAC2008-Z-2008} to give an analytic nonrecursive 
expression of all the admissible trajectories. A deeper insight into the features of these subspaces, particularly focused on the dimensionality issue,
is the object of this note.
\end{abstract}
%
%%%%%%%%%%%%%%%%%%%%%%%%%%%%%%%%%%%%%%%%%%%%%%%%%%%%%%%%%%%%%%%%%%%%%%%%%%%%%%%%
%
\section{Main Content}
\label{sect_insight}
In \cite{TAC2008-Z-2008}, the structural invariant subspaces $\gV_1$ and $\gV_2$, defined by
(17) and (18) respectively, are used in (19), that is the analytic nonrecursive expression of
the set of the admissible  solutions of the discrete-time singular Hamiltonian  system (10)
over the time interval $0\,{\le}\,k\,{\le}\,k_f\,{-}\,1$.

In this note it will be shown that the dimension of the subspace  $\gV_2$ may be  lower
than $n$, where $n$ denotes the dimension of the state space of the original system as defined in (1),\,(2).
In particular, the possible loss of dimension of $\gV_2$ (or, equivalently, the possible loss 
of rank of the matrix $V_2$ defined by (18)) depends on the properties of the original system
(1),\,(2) (under assumptions $\gA.1$--$\gA.4$).
\par
Let us consider system~(1),\,(2) and perform the similarity 
transformation $T\,{=}\,\left[T_1\ T_2\right]$, where ${\rm im}\,T_1\,{=}\,\gR$, the
reachable subspace of $(A,B)$. With respect to the new basis, 
\begin{eqnarray*}
A\ns&\ns=\ns&\ns
\left[\begin{array}{cc}
A_c & A_{cu}\\
O   & A_u
\end{array}\right],\
B=
\left[\begin{array}{c}
B_c\\
O  
\end{array}\right],\\
C\ns&\ns=\ns&\ns\left[\begin{array}{cc}
C_c & C_u
\end{array}\right],\quad
D=D.
\end{eqnarray*}
Moreover, the solution $P_+$ of the Riccati equation~(11),\,(12), partitioned accordingly, is
\[
P_+=
\left[\begin{array}{cc}
P_c & P_{cu}\\
P_{cu}^\top & P_u
\end{array}\right],
\]
where $P_c$ is the stabilizing solution of the Riccati equation restricted to the sole 
reachable part of the original system: i.e.,
\[
P_c = A_c^\top P_cA_c+C_c^\top C_c
-(A_c^\top P_cB_c+C_c^\top D)(D^\top D+B_c^\top P_cB_c)^{-1}(B_c^\top P_cA_c+D^\top C_c),
\]
with
\[
D^\top D+B_c^\top P_cB_c>0.
\]
The stabilizing feedback is partitioned as
\[
-K_+=\left[\begin{array}{cc}
K_c & K_u
\end{array}\right].
\]
Similarly, the solution $W$ of the discrete Lyapunov equation has the structure
\[
W=
\left[\begin{array}{cc}
W_c & O\\
O & O
\end{array}\right],
\]
where $W_c$ is the solution of the discrete Lyapunov equation restricted to the sole 
reachable part of the original system: i.e.,
\[
(A_c+B_cK_c)W_c(A_c+B_cK_c)^\top+B_c(D^\top D+B_c^\top P_cB_c)^{-1}B_c^\top = W_c.
%\label{eq_lyaprestr}
\]
Simple algebraic manipulations, where these partitions are taken into account, yield
the following structure for the matrix $V_2$:
\[
%\label{eq_V2}
V_2=\left[\begin{array}{cc}
W_c(A_c+B_cK_c)^\top & O\\
O & O\\
\star & O\\
\star & -A_u^\top\\
\star & O
\end{array}
\right],
\]
where the symbol $\star$ denotes a possibly nonzero submatrix. The structure pointed out
in the partitioned matrix $V_2$ shows that the rank of $V_2$ may be lower than $n$. This circumstance 
occurs, for instance, if $A_u$ has a zero row, like in the illustrative example considered in 
the following section. 

It is worth noting that, by contrast, the subspace 
\[ 
\bar\gV_2={\rm im}\,\left[\begin{array}{c} W \\ P_+W-I \end{array}\right]\,,
\]
that is used in (20) of \cite{TAC2008-Z-2008} in order to express the sole state and costate trajectories
over the time interval $0\,{\le}\,k\,{\le}\,k_f$, has dimension $n$. 

In fact, the corresponding  partitioned matrix is
\[
\bar V_2=\left[\begin{array}{cc}
W_c & O\\
O   & O\\
\star & O\\
\star & -I
\end{array}
\right].
\]
The rank of $\bar V_2$ is $n$, since the symmetric positive definite $W_c$, being the solution of 
the restricted Lyapunov equation above, has the same rank of the controllability Gramian of the pair
$(A_c\,{+}\,B_cK_c,B_c)$, which is completely controllable by construction.

%%%%%%%%%%%%%%%%%%%%%%%%%%%%%%%%%%%%%%%%%%%%%%%%%%%%%%%%%%%%%%%%%%%%%%%%%%%%%%%%%%%%%%%

\section{An Illustrative Example}
\label{sect_example}
This section presents a numerical example where the rank of matrix $V_2$ 
is lower than the dynamic order $n$ of the original system, while $n$ is the rank of matrix
$\bar V_2$. The variables are displayed in scaled fixed point format with 
five digits, although computations are made in floating point precision.
Consider system~(1),\,(2) in \cite{TAC2008-Z-2008}, with
\begin{eqnarray*}
A\ns&\ns=\ns&\ns \left[\begin{array}{cccc}
    0.3	&   -0.4	&    0.5	&    0.6\\
    0.1	&    0.2	&    0.1	&    0.1\\
    0	&    0		&    0.5	&    0\\
    0	&    0		&    0		&    0
\end{array}\right],\quad
B=\left[\begin{array}{cc}
     1	&     0.2\\
     2	&     3\\
     0	&     0\\
     0 	&     0
\end{array}\right],\label{eq_AB}\\
C\ns&\ns=\ns&\ns \left[\begin{array}{cccc}
      1	&     2	&     3	&	4\\	
      2	&     1	&     5	&	6
\end{array}\right],\hspace{21mm}
D=\left[\begin{array}{cc}
     10	&     0\\
     0  &     0
\end{array}\right]\label{eq_CD},
\end{eqnarray*}
By pursuing the procedure illustrated in \cite{TAC2008-Z-2008}, one gets, in particular
\[
V_2= \left[\begin{array}{cccc}
    0.3661 & -0.4314  &      0    &    0\\
   -0.7323 &  0.8629  &      0    &    0\\
         0 &       0  &      0    &    0\\
         0 &       0  &      0    &    0\\
   -0.0000 &  0.0000  &      0    &    0\\
    0.0000 & -0.0000  &      0    &    0\\
    1.8443 & -1.1885  &-0.5000    &    0\\
   -0.0000 &  0.0000  &      0    &    0\\
    0.1098 & -0.1294  &      0    &    0\\
   -0.2088 &  0.4374  &      0    &    0
\end{array}\right],
\]
and
\[
\bar V_2= \left[\begin{array}{cccc}
    0.4708 & -0.5165 &       0 &       0\\
   -0.5165 &  1.1828 &       0 &       0\\
         0 &       0 &       0 &       0\\
         0 &       0 &       0 &       0\\
   -0.0000 &  0.0000 &       0 &       0\\
    0.0000 & -0.0000 &       0 &       0\\
    3.6885 & -2.3770 & -1.0000 &       0\\
    2.7295 &  0.5411 &       0 & -1.0000
\end{array}\right].
\]
%
%%%%%%%%%%%%%%%%%%%%%%%%%%%%%%%%%%%%%%%%%%%%%%%%%%%%%%%%%%%%%%%%%%%%%%%%%%%%%%%%%%%%%%%%
%
\section{Conclusions}
\label{sect_conclusions}
In this note, it has been shown that the dimension of the structural invariant subspace 
$\gV_2$ may be lower than the dynamic order $n$ of the original system. The result has
been illustrated by a numerical example. This is the reason why the only-if part of the proof 
of Theorem~1 in \cite{TAC2008-Z-2008} does not rely on a dimensionality count, but on
the maximality of the subspace $\gV_2$ (and of the subspace $\gV_1$). In fact, maximality 
follow from Property~2 (and Property~1, respectively), according to \cite[Section~5.4]{Ionescu-OW-1999}.
%
%%%%%%%%%%%%%%%%%%%%%%%%%%%%%%%%%%%%%%%%%%%%%%%%%%%%%%%%%%%%%%%%%%%%%%%%%%%%%%%%%%%%%%%%
%
\bibliographystyle{IEEEtran}
\bibliography{remarksbib}
\end{document}